\documentclass[11pt]{article}
\usepackage[applemac]{inputenc}
\usepackage{cite}
\usepackage{fullpage}
\usepackage{amssymb}
\usepackage{amsmath}
\usepackage{graphicx}
\usepackage{enumerate}

\newcommand{\str}{\ensuremath{\mathrm{str}}} 
\newcommand{\strdepth}{\ensuremath{\mathrm{strdepth}}} 
 
\newcommand{\locus}{\ensuremath{\mathrm{locus}}} 
 
\newcommand{\lab}{\ensuremath{\mathrm{label}}} 
\newcommand{\order}{\ensuremath{\mathrm{order}}} 
\newcommand{\srr}{substring range reporting}
\newcommand{\parent}{\ensuremath{\mathrm{parent}}} 

\newcommand{\report}{\ensuremath{\textsc{report}}}
\newcommand{\pmi}{indexing substrings with intervals}
\newcommand{\Pmi}{Indexing substrings with intervals}
\newcommand{\gaps}{indexing substrings with gaps}
\newcommand{\Gaps}{Indexing substrings with gaps}

\newcommand{\sort}{\ensuremath{\mathrm{sort}}} 

\newcommand{\occ}{\ensuremath{\mathrm{occ}}} 

\newtheorem{lemma}{Lemma}
\newtheorem{theorem}{Theorem}

\title{Substring Range Reporting\footnote{An extended abstract of this paper appeared at the 22nd Conference on Combinatorial Pattern Matching.}}
\author{Philip Bille \\ \texttt{phbi@imm.dtu.dk} \and Inge Li G{\o}rtz \\ \texttt{ilg@imm.dtu.dk}}

\begin{document}
\maketitle

\begin{abstract}
We revisit various string indexing problems with range reporting features, namely, position-restricted substring searching, indexing substrings with gaps, and indexing substrings with intervals. We obtain the following main results.
\begin{itemize}
\item We give efficient reductions for each of the above problems to a new problem, which we call \emph{substring range reporting}. Hence, we unify the previous work by showing that we may restrict our attention to a single problem rather than studying each of the above problems individually. 
\item We show how to solve substring range reporting with optimal query time and little space. Combined with our reductions this leads to significantly improved time-space trade-offs for the above problems. In particular, for each problem we obtain the first solutions with optimal time query and $O(n\log^{O(1)} n)$ space, where $n$ is the length of the indexed string. 
\item We show that our techniques for substring range reporting generalize to \emph{substring range counting} and \emph{substring range emptiness} variants. We also obtain non-trivial time-space trade-offs for these problems.
\end{itemize}
Our bounds for substring range reporting are based on a novel combination of suffix trees and range reporting data structures. The reductions are simple and general and may apply to other combinations of string indexing with range reporting.
\end{abstract}

\section{Introduction}
Given a string $S$ of length $n$ the \emph{string indexing problem} is to preprocess  $S$ into a compact representation that efficiently supports \emph{substring queries}, that is, given another string $P$ of length $m$ report all occurrences of substrings in $S$ that match $P$. Combining the classic suffix tree data structure~\cite{Gusfield1997} with perfect hashing~\cite{FKS1984} leads to an optimal time-space trade-off for string indexing, i.e., an $O(n)$ space representation that supports queries in $O(m + \occ)$ time, where $\occ$ is the number of occurrences of $P$ in $S$.

In recent years, several extensions of string indexing problems that add \emph{range reporting} features have been proposed. For instance, M\"{a}kinen and Navarro proposed the \emph{position-restricted substring searching problem}~\cite{MN2006,MN2007}. Here, queries take an additional range $[a,b]$ of positions in $S$ and the goal is to report the occurrences of $P$ within $S[a,b]$. For such extensions of string indexing no optimal time-space trade-off is known. For instance, for position-restricted substring searching one can either get $O(n\log^\varepsilon n)$ space (for any constant $\varepsilon > 0$)   and $O(m + \log \log n + \occ)$ query time or $O(n^{1+\varepsilon})$ space with $O(m + \occ)$ query time~\cite{MN2006,MN2007, CIKRW2008}. Hence, removing the $\log \log n$ term in the query comes at the cost of significantly increasing the space.

In this paper, we revisit a number string indexing problems with range reporting features, namely \emph{position-restricted substring searching}, \emph{indexing substrings with gaps}, and \emph{indexing substrings with intervals}. We achieve the following results. 
\begin{itemize}
\item We give efficient reductions for each of the above problems to a new problem, which we call \emph{substring range reporting}. Hence, we unify the previous work by showing that we may restrict our attention to a single problem rather than studying each of the above problems individually. 

\item We show how to solve substring range reporting with optimal query time and little space. Combined with our reductions this leads to significantly improved time-space trade-offs for all of the above problems. For instance, we show how to solve position-restricted substring searching in $O(n\log^{\varepsilon} n)$ space and $O(m + \occ)$ query time. 

\item We show that our techniques for substring range reporting generalize to \emph{substring range counting} and \emph{substring range emptiness} variants. We also obtain non-trivial time-space trade-offs for these problems.

\end{itemize}
Our bounds for substring range reporting are based on a novel combination of suffix trees and range reporting data structures. The reductions are simple and general and may apply to other combinations of string indexing with range reporting.

\subsection{Substring Range Reporting}\label{sec:srrintro}
Let $S$ be a string where each position is associated with a integer
value in the range $[0,u]$. The integer associated with position $i$ in $S$ is the
\emph{label} of position $i$, denoted $\lab(i)$, and we call $S$ a \emph{labeled string}. Given a labeled string $S$, the \emph{substring range reporting problem} is to compactly represent $S$ while supporting
\emph{substring range reporting queries}, that is, given a string $P$
and a pair of integers $a$ and $b$, $0 \leq a \leq b \leq u$, report all starting positions in $S$ that match $P$ and whose labels are in the range $[a, b]$.

We assume a standard unit-cost RAM model with word size $w$ and a standard instruction set including arithmetic operations, bitwise boolean operations, and shifts. We assume that a label can be stored in a constant number of words and therefore $w = \Theta(\log u)$.  The space complexity is the number of words used by the algorithm. All bounds mentioned in this paper are valid in this model of computation.

To solve substring range reporting a basic approach is to combine a suffix tree with a 2D range reporting data structure. A query for a pattern $P$ and range $[a,b]$ consists of a search in the suffix tree and then a 2D range reporting query with $[a,b]$ and the lexicographic range of suffixes defined $P$. This is essentially the overall approach used in the known solutions for position-restricted substring searching~\cite{MN2006,MN2007, CIKRW2008,CIKRW2010,YHW2011, BMMM2009}, which is a special case of substring range reporting (see the next section). 

Depending on the choice of the 2D range reporting data structure this approach leads to different trade-offs. In particular, if we plug in the 2D range reporting data structure of Alstrup et al.~\cite{ABR2000}, we get a solution with $O(n\log^\varepsilon n)$ space and $O(m + \log \log u + \occ)$ query time (see M\"{a}kinen and Navarro~\cite{MN2006,MN2007}). The $\log \log u$ term in the query time is from the range reporting query. Alternatively, if we use a fast data structure for the range successor problem~\cite{CIKRW2008, YHW2011} to do the range reporting, we get optimal $O(m + \occ)$ query time but increase the space to at least $\Omega(n^{1+\varepsilon})$. Indeed, since any 2D range reporting data structure with $O(n \log^{O(1)} n)$ space must use $\Omega(\log \log u)$ query time~\cite{PT2006}, we cannot hope to avoid this blowup in space with this approach. 

Our first main contribution is a new and simple technique that overcomes the inherent problem of the previous approach. We show the following result.
\begin{theorem}\label{thm:main}
Let $S$ be a labeled string of length $n$ with labels in the range $[0,u]$. For any constants $\varepsilon, \delta >0$, we can solve substring range reporting using $O(n(\log^\varepsilon n + \log \log u))$ space, $O(n(\log n + \log^\delta u))$ expected preprocessing time, and $O(m + \occ)$ query time, for a pattern string of length $m$. 
\end{theorem}
Compared to the previous results we achieve optimal query time with an additional $O(n \log \log u)$ term in the space. For the applications considered here, we have that $u = O(n)$ and therefore the space bound simplifies to $O(n (\log^\varepsilon n + \log \log u)) = O(n \log^\varepsilon n)$. Hence, in this case there is no asymptotic space overhead.

The key idea to obtain Theorem~\ref{thm:main} is a new and simple combination of suffix trees with multiple range reporting data structures for both 1 and 2 dimensions. Our solution handles queries differently depending on the length of the input pattern such that the overall query is optimized accordingly.

Interestingly, the idea of using different query algorithms depending on the length of the pattern is closely related to the concept of \emph{filtering search} introduced for the standard range reporting problem by Chazelle as early as  1986~\cite{Chazelle1986}. Our new results show that this idea is also useful in combinatorial pattern matching.

Finally, we also consider \emph{substring range counting} and \emph{substring range emptiness} variants. Here, the goal is to count the number of occurrences in the range and to determine whether or not the range is empty, respectively. Similar to substring range reporting, these problems can also be solved in a straightforward way by combining a suffix with a 2D range counting or emptiness data structure. We show how to extend our techniques to obtain improved time-space trade-offs for both of these problems.

\subsection{Applications} 
Our second main contribution is to show that substring range reporting actually captures several  other string indexing problems. In particular, we show how to reduce the following problems to substring range reporting.

\begin{itemize}
\item[$\bullet$] \emph{Position-restricted substring searching:} Given a string $S$ of length $n$, construct a data structure supporting the following query: Given a string $P$ and query interval $[a,b]$, with $1 \leq a \leq b\leq n$, return the positions of substrings in $S$ matching $P$ whose positions are in the interval $[a,b]$.

\item[$\bullet$] \emph{\Pmi:} Given a string $S$ of length $n$, and a set of intervals $\pi =\{[s_1,f_1],[s_2,f_2],\ldots, [s_{|\pi|},f_{|\pi|}]\}$ such that $s_i,f_i \in [1,n]$ and $s_i \leq f_i$, for all $1 \leq i \leq |\pi |$, construct a data structure supporting the following query: Given a string $P$ and query interval $[a,b]$, with $1 \leq a \leq b\leq n$, return the positions of substrings in $S$ matching $P$ whose positions are in $[a,b]$ \emph{and} in one of the intervals in $\pi$.

\item[$\bullet$] \emph{\Gaps:}
Given a string $S$ of length $n$ and an integer $d$, the problem is to construct a data structure supporting the following query: Given two strings $P_1$ and $P_2$ return all positions of substrings in $S$ matching $P_1 \circ \star^d \circ P_2$. Here $\circ$ denotes concatenation and $\star$ is a wildcard matching all characters in the alphabet. 
\end{itemize}

\paragraph{Previous results} Let $m$ be the length of $P$. M\"akinen and Navarro~\cite{MN2006, MN2007} introduced the position-restricted substring searching problem. Their fastest solution uses $O(n \log ^{\varepsilon} n)$ space, $O(n \log n)$ expected preprocessing time, and $O(m + \log\log n+ \occ)$ query time. Crochemore~et~al.~\cite{CIKRW2008} proposed another solution using $O(n^{1+\varepsilon})$ space, $O(n^{1+\varepsilon})$ preprocessing time, and $O(m + \occ)$ query time (see also Section~\ref{sec:srrintro}). Using techniques from range non-overlapping indexing~\cite{CP2009} it is possible to improve these bounds for small alphabet sizes~\cite{Porat2011}. Several succinct versions of the problem have also been proposed~\cite{MN2006, MN2007,YHW2011, BMMM2009}. All of these have significantly worse query time since they require superconstant time per reported occurrence. Finally, Crochemore et al.~\cite{CIR2008} studied a restricted version of the problem with $a = 1$ or $b = n$.

For the \pmi\ problem, Crochemore et al.~\cite{CIKRW2008,CIKRW2010} gave a solution with $O(n \log^2 n)$ space, $O(|\pi|+ n \log^3 n)$ expected preprocessing time, and $O(m + \log\log n+  \occ)$ query time. They also showed how to achieve $O(n^{1+\varepsilon})$ space, $O(n^{1+\varepsilon}+ |\pi |)$ preprocessing time, and $O(m +  \occ)$ query time. Several papers~\cite{ACIKZ2008,IR2008,JLW2009} have studied the property matching problem, which is similar to the \pmi\ problem, but where both start and end point of the match must be in the same interval.

Iliopoulos and Rahman~\cite{IR2009} studied the problem of \gaps. They gave a solution using $O(n \log^{\varepsilon} n)$ space, $O(n \log n)$ expected preprocessing time, and $O(m + \log\!\log n + \occ)$ query time, where $m$ is the length of the two 
input strings. Crochemore and Tischler recently proposed a variant of the problem~\cite{CT2010}.

\paragraph{Our results} We reduce position-restricted substring searching, \pmi, and \gaps\ to \srr. Applying Theorem~\ref{thm:main} 
with our new reductions, we get the following result.
\begin{theorem}\label{thm:app}
Let $S$ be a string of length $n$ and let $m$ be the length of the query. For any constant $\varepsilon>0$, we can solve
\begin{enumerate}[(i)]
\item\label{thm:app:prss} Position-restricted substring searching using $O(n \log^{\varepsilon} n)$ space, $O(n \log n)$ expected preprocessing time, and $O(m + \occ)$ query time. 
\item\label{thm:app:pmi} \Pmi\ using $O(n \log^{\varepsilon} n)$ space, $O(|\pi|+ n\log n)$ expected preprocessing time, and $O(m + \occ)$ query time. 
\item\label{thm:app:gaps} \Gaps\ using $O(n\log^{\varepsilon} n)$ space, $O(n\log n)$ expected preprocessing time, and $O(m +  \occ)$ query time ($m$ is the size of the two input strings).
\end{enumerate}
\end{theorem}
This improves the best known time-space trade-offs for all three problems, that all suffer from the trade-off inherent in 2D range reporting.

The reductions are simple and general and may apply to other combinations of string indexing with range reporting.

\section{Basic Concepts}\label{sec:preliminaries}
\subsection{Strings and  Suffix Trees}
Throughout the section we will let $S$ be a labeled string of length $|S|=n$
with labels in $[0,u]$. We denote the character at
position $i$ by $S[i]$ and the substrings from position $i$ to $j$ by
$S[i,j]$. The substrings $S[1,j]$ and $S[i, n]$ are the
\emph{prefixes} and \emph{suffixes} of $S$, respectively. The \emph{reverse} of $S$ is $S^R$. We denote the label of
position $i$ by $\lab_S(i)$. The \emph{order} of suffix $S[i, n]$,
denoted $\order_S(i)$, is the lexicographic order of $S[i, n]$ among
the suffixes of $S$.

The \emph{suffix tree} for $S$, denoted $T_S$, is the compacted trie
storing all suffixes of $S$~\cite{Gusfield1997}. The \emph{depth} of a node $v$ in $T_S$ is the number of edges on the path from $v$ to the root. Each of the edges in $T_S$ is associated with some substring of $S$. The children of each node are sorted from left to right in increasing alphabetic order of the first character of the substring associated with the edge leading to them. The concatenation of substrings from the root to $v$ is denoted
$\str_S(v)$. The \emph{string depth} of $v$, denoted
$\strdepth_S(v)$, is the length of $\str_S(v)$. The \emph{locus} of a string $P$, denoted $\locus_S(P)$, is the minimum depth node $v$ such that $P$ is a prefix of $\str_S(v)$. If $P$ is not a prefix of a substring in $S$ we define $\locus_S(P)$ to be $\bot$.

Each leaf $\ell$ in $T_S$ uniquely corresponds to a suffix in $S$, namely, the suffix $\str_S(\ell)$.  Hence, we will use $\lab_S(\ell)$ and $\order_S(\ell)$
to refer to the label and order of the corresponding suffix. For an
internal node $v$ we extend the notation such that
\begin{align*}
\lab_S(v) &= \{\lab_S(\ell) \mid \ell \text{ is a descendant leaf of v}\} \\
\order_S(v) &= \{\order_S(\ell) \mid \ell \text{ is a descendant leaf of v}\}.
\end{align*}
Since children of a node are sorted, the left to right order of the
leaves in $T_S$ corresponds to the lexicographic order of the suffixes
of $S$. Hence, for any node $v$, $\order_S(v)$ is an interval. We
denote the left and right endpoints of this interval by $l_v$ and
$r_v$. When the underlying string $S$ is clear from the context we will often drop the subscript $_S$ for brevity.

The suffix tree for $S$ uses $O(n)$ space and can be constructed in $O(\sort(n))$ time, where $\sort(n)$ is the time for sorting $n$ values in the model of computation~\cite{CFM2000}. We only need a standard comparison-based $O(n\log n)$ suffix tree construction in our results. Let $P$ be a string of length $m$. If $\locus_S(P) = \bot$ then $P$ does not occur as a substring in $S$. Otherwise, the substrings in $S$ that match $P$ are the suffixes in $\order_S(\locus_S(P))$. Hence, we can compute all occurrences of $P$ in $S$ by traversing the suffix tree from the root to $\locus_S(P)$ and then report all suffixes stored in the subtree. Using perfect hashing~\cite{FKS1984} to represent the outgoing edges of each node in $T_S$ we achieve an $O(n)$ solution to string indexing that supports queries in $O(m + \occ)$ time (here $\occ$ is the total number of occurrences of $P$ in $S$).

\subsection{Range Reporting}
Let $X \subseteq \{0, \ldots, u\}^d$ be a set of points in a d-dimensional grid.  The \emph{range reporting problem} in $d$-dimensions is to compactly represent $X$ while supporting \emph{range reporting queries}, that is, given a rectangle $R = [a_1, b_1] \times \cdots \times [a_d, b_d]$ report all points in the set $R
\cap X$.  We use the following results for range reporting in $1$ and $2$ dimensions.

\begin{lemma}[Alstrup et al.~\cite{ABR2001}, Mortensen et
  al.~\cite{MPP2005}]\label{lem:1DRR} 
  For a set of $n$ points in $[0, u]$ and any constant $\gamma >0$, we can solve 1D range reporting using $O(n)$ space, $O(n\log^\gamma u)$ expected preprocessing time and $O(1 + \occ)$ query time.
  \end{lemma}

\begin{lemma}[Alstrup et al.~\cite{ABR2000}]\label{lem:2DRR}
  For a set of $n$ points in $[0, u] \times [0,u]$ and any constant $\varepsilon > 0$, we can solve 2D range reporting using $O(n\log^\varepsilon n)$  space, $O(n\log n)$ expected preprocessing time, and $O(\log \log u + \occ)$ query time. 
\end{lemma}

\section{Substring Range Reporting}\label{sec:ds}
We now show Theorem~\ref{thm:main}. Recall that $S$ is a labeled string of length $n$ with labels from $[0, u]$. 

\subsection{The Data Structure}
Our substring range reporting data structure consists of the
following components.
\begin{itemize}
\item The suffix tree $T_S$ for $S$. For each node $v$ in $T_S$ we
  also store $l_v$ and $r_v$.  We partition $T_S$ into a \emph{top
    tree} and a number of \emph{bottom trees}. The top tree consists
  of all nodes in $T_S$ whose string depth is at most $\log \log u$ and all
  their children. The trees induced by the remaining nodes are the forest
  of bottom trees.
\item A 2D range reporting data structure on the set of points
  $\{(\order_S(i), \lab_S(i)) \mid i \in \{1, \ldots, n\}\}$. 
\item For each node $v$ in the top tree, a 1D range reporting data
  structure on the set $\{\lab_S(i) \mid i \in \order_S(v)\}$.
\end{itemize}
We analyze the space and preprocessing time for the data structure. We use the range reporting data structures  from Lemmas~\ref{lem:1DRR} and \ref{lem:2DRR}. 
The space for the suffix tree is $O(n)$ and the space for the 2D range reporting data structure is $O(n \log^\varepsilon n)$, for any constant $\varepsilon > 0$. We bound the space for the (potentially $\Omega(n)$) 1D range reporting data structures stored for the top tree. Let $V_d$ be the set of nodes in the top tree with
depth $d$. Since the sets $\order_S(v)$, $v \in V_d$, partition the set of descendant leaves of nodes in $V_d$, the total size of these sets is as most $n$. Hence, the total size of the 1D range reporting data structures for the nodes in $V_d$ is therefore $O(n)$. Since there are at most $\log \log u + 1$ levels in the top tree, the space for all 1D range reporting data structures is $O(n \log \log u)$. Hence, the total space for the data structure is $O(n (\log^\varepsilon n + \log \log u))$.

We can construct the suffix tree in $O(\sort(n))$ time and the 2D range reporting data structure in $O(n \log n)$ expected time. For any constant $\gamma > 0$, the expected preprocessing time for all 1D range reporting data structures is  
$$
O\left(\sum_{v \text{ in top tree}} |\order_S(v)| \log^{\gamma} u
\right) = O(n \log \log u \log^{\gamma} u) = O(n \log^{2\gamma} u). 
$$ 
Setting $\delta = 2\gamma$ we use $O(n (\log n + \log^\delta u))$ expected  
preprocessing time in total.

\subsection{Substring Range Queries}
Let $P$ be a string of length $m$, and let $a$ and $b$ be a pair of
integers, $0 \leq a \leq b \leq u$. To answer a substring range query
we want to compute the set of starting positions for $P$ whose labels
are in $[a, b]$. First, we compute the node $v = \locus_S(P)$. If $v = \bot$ then $P$ is not a substring of $S$, and we return the empty set. Otherwise, we compute the set of descendant leaves of $v$ with labels in $[a,b]$. There are two
cases to consider.
\begin{enumerate}[(i)]
\item If $v$ is in the top tree we query the 1D range reporting data
  structure for $v$ with the interval $[a, b]$.
\item If $v$ is in a bottom tree, we query the 2D range reporting data
  with the rectangle $[l_v, r_v] \times [a, b]$.
\end{enumerate}
Given the points returned by the range reporting data structures, we output the corresponding starting positions of the corresponding suffixes. From the definition of the data structure it follows that these are precisely the occurrences of $P$ within the range $[a,b]$. Next consider the time complexity. We find $\locus_S(P)$ in $O(m)$ time (see Section~\ref{sec:preliminaries}). In case (i) we use $O(1
+ \occ)$ time to compute the result by Lemma~\ref{lem:1DRR}. Hence,
the total time for a substring range query for case (i) is $O(m +
\occ)$.  In case (ii) we use $O(\log \log u + \occ)$ time to compute the
result by Lemma~\ref{lem:2DRR}. We have that $v=
\locus_S(P)$ is in a bottom tree and therefore $m \geq
\strdepth(\parent(\locus_S(v))) > \log \log u$. Hence, the total time to
answer a substring range query in case (ii) is $O(m + \log \log u +
\occ) = O(m + \occ)$. Thus, in both cases we use $O(m + \occ)$
time. 

Summing up, our solution uses $O(n (\log^\varepsilon n + \log \log u)$ space, $O(n (\log n + \log^\delta u))$ expected preprocessing time, and $O(m + \occ)$ query time. This completes the proof of Theorem~\ref{thm:main}. 

\section{Applications}\label{sec:applications}
In this section we show how to improve the results for the three problems position-restricted substring searching, \pmi, and indexing gapped substrings, using our data structure for \srr. Let $\report_S(P,a,b)$ denote a \srr\ query on string $S$ with parameters $P$, $a$, and $b$.

\subsection{Position-Restricted Substring Searching}
We can reduce position-restricted substring searching to \srr\ by setting $\lab(i)=i$ for all $i=1,\ldots,n$. To answer a query we return the result of the substring range query $\report_S(P,a,b)$. Since each label  is equal to the position, it follows that the solution to the \srr\ instance immediately gives a solution to position-restricted substring searching. 
Applying Theorem~\ref{thm:main} with $u=n$, this proves Theorem~\ref{thm:app}(\ref{thm:app:prss}).


\subsection{Indexing Substrings with Intervals}
We can reduce \pmi\ to \srr\ by setting $$
\lab(i) = 
\begin{cases}
i & \textrm{if } i \in \varphi  \textrm{ for some } \varphi \in \pi,\\
0 & \textrm{otherwise}.
\end{cases}
$$
To answer a query 
we return the result of the \srr\ query $\report_S(P,a,b)$.
Let $I$ be the solution to the \pmi\ instance and let $I'$ be the solution to the \srr\ instance derived by the above reduction. Then $i \in I \Leftrightarrow i \in I'$. 
 
 To prove this assume $i \in I$. Then $i \in \varphi$ for some  $\varphi \in \pi$ and $i \in [a,b].$ 
 From $i \in \varphi$ and the definition of $\lab(i)$ it follows that $\lab(i) = i$. Thus, $\lab(i) = i \in [a,b]$ and thus $i \in I'$. Assume $i \in I'$. Then $\lab(i) \in [a,b]$. Since $a > 0$ also $\lab(i) > 0$, and it follows that $\lab(i) = i$. By the reduction this means that $i \in  \varphi$ for some  $\varphi \in \pi$. Since $i =\lab(i)$, we have $i \in [a,b]$ and therefore $i \in I$.

We can construct the labeling in $O(n + |\pi|)$ if the intervals are sorted by startpoint or endpoint. Otherwise additional time for sorting is needed. A similar approach is used in the solution by Crochemore et al.~\cite{CIKRW2008}.
 
Applying Theorem~\ref{thm:main} with $u=n$, this proves Theorem~\ref{thm:app}(\ref{thm:app:pmi}).


\subsection{Indexing Substrings with Gaps}
\begin{figure}[t]
\centering
\includegraphics[scale=.5]{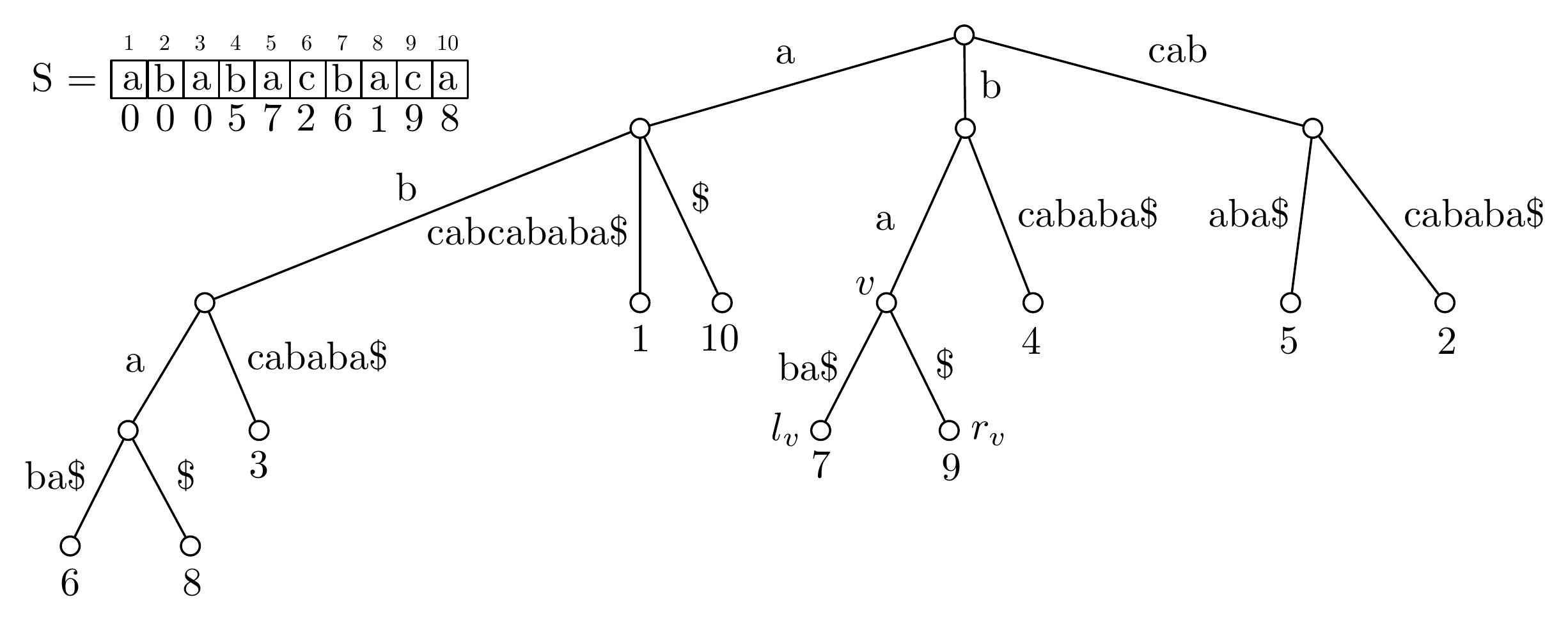}
\caption{A string $S$, the labeling for $d = 2$ (below the string), and the suffix tree of $T_{S^R}$. Given a query $P_1= \text{ab}$ and $P_2=\text{bac}$ we find $v=\locus_{S^R}(\mathrm{ba})$ (marked in the suffix tree). We have $l_v=6$ and $r_v=7$ from the left-to-right-order in the $T_{S^R}$. The \srr\ query $\report_s(P_2, 6,7)$ returns 7. Hence, we report the occurrence at position  $7-2-2=3$.}
\label{fig:ipg}
\end{figure}

We can reduce the \gaps\ problem to \srr\ as follows. Construct the suffix tree of the reverse of $S$, i.e., the suffix tree $T_{S^R}$ for $S^R$. For each node $v$ in $T_{S^R}$ also store $l_v$ and $r_v$. Set $$\lab_S(i) = 
		\begin{cases} \order_{S^R}(n - i+d +2) & \textrm{for } i \geq d +2,  \\ 
				  0 &\textrm{otherwise}.
		\end{cases}
		$$
To answer a query find the locus node $v$ of $P_1^R$ in $T_{S^R}$. Then use the \srr\ data structure to return all positions of substrings in $S$ matching $P_2$ whose labels are in the range $[l_v,r_v]$. For each position $i$ returned by $\report_S(P_2,l_v,r_v)$, return $i-|P_1|-d$. See~Fig.~\ref{fig:ipg} for an example.

\paragraph{Correctness of the reduction} We will now show that the reduction is correct.  Let $I$ be the solution to the \gaps\ instance and let $I'$ be the solution to the \srr\ instance derived by the above reduction. We will show $i \in I \Leftrightarrow i \in I'$. Let $m_i=|P_i|$ for $i=1,2$. 

If $i \in I$ then there is 
an occurrence of $P_1$ at position $i$ in $S$ and an occurrence of $P_2$ at position $i'=i+m_1+d$ in $S$.  It follows directly, that there is an occurrence of $P_1^R$  at position $i''=n-(i+m_1)+2$ in $S^R$.
By definition, $$\lab_S(i') =\lab_S(i+m_1+d) =  \order_{S^R}(n - (i+m_1+d)+d +2) = \order_{S^R}(i''),$$
where the second equality follows from the fact that $i+m_1+d\geq d+2$. Since there is an occurrence of $P_1^R$  at position $i''$ in $S^R$, we have $\lab_S(i')=\order_{S^R}(i'') \in \order_{S^R}(\locus_{S^R}(P_1^R)).$ Thus, $\lab_S(i') \in [l_v,r_v]$, and since there is an occurrence of $P_2$ at position $i'$ in $S$, we have $i' - m_1-d= i \in I'$.

If $i \in I'$ then there is an occurrence of $P_2$ at position $i'=i+m_1+d$ with $\lab(i')$ in the range $[l_v,r_v]$, where 
$v=\locus_{S^R}(P_1^R)$. 
 We need to show that this implies that there is an occurrence of $P_1$ at position $i$ in $S$. By definition, 
$$ \lab_S(i')= \order_{S^R}(n - i'+d +2)
=\order_{S^R}(n - i-m_1 +2).
$$
Let $i''=n - i-m_1 +2$. Since $\order_{S^R}(i'') =\lab_S(i') \in [l_v,r_v]$, there is an occurrence of $P_1^R$ at position $i''$ in $S^R$. It follows directly, that there is an occurrence of $P_1$ at position $n-i''-m_1+2=n-(n - i-m_1 +2)-m_1+2=i$ in $S$. Therefore, $i \in I$.

\paragraph{Complexity} 
Construction of the suffix tree $T_{S^R}$ takes time $O(n\log n)$ and the labeling can be constructed in time $O(n)$.  Both use space $O(n)$. 
 It takes $O(m_1)$ time to find the locus nodes of $P_1^R$ in $T_{S^R}$. The \srr\ query takes time $O(m_2 + \occ)$. Thus the total query time is $O(m+\occ)$.

Applying Theorem~\ref{thm:main} with $u=n$, this completes the proof of Theorem~\ref{thm:app}(iii). 

\section{Substring Range Counting and Emptiness}\label{sec:extensions}
We now show how to apply our techniques to \emph{substring range counting} and \emph{substring range emptiness}. Analogous to substring range reporting, the goal is here to count the number of occurrences in the range and to determine whether or not the range is empty, respectively. A straightforward way to solve these problems is to combine a suffix tree with a 2D range counting data structure and a 2D range emptiness data structure, respectively. Using the techniques from Section~\ref{sec:ds} we show how to significantly improve the bounds of this approach in both cases. We note that by the reductions in Section~\ref{sec:applications} the bounds for substring range counting and substring range emptiness also immediately imply results for counting and emptiness versions of position-restricted substring searching, indexing substrings with intervals, and indexing substrings with gaps.

\subsection{Preliminaries}
Let $X \subseteq \{0, \ldots, u\}$ be a set of points in a $d$-dimensional grid. Given a query rectangle $R = [a_1, b_1] \times \cdots \times [a_d, b_d]$, a \emph{range counting query} computes $|R \cap X|$, and a \emph{range emptiness query} computes if $R \cap X = \emptyset$. Given $X$ the \emph{range counting problem} and the \emph{range emptiness problem} is to compactly represent $X$, while supporting range counting queries and range emptiness queries, respectively. Note that any solution for range reporting or range counting implies a solution for range emptiness with the same complexity (ignoring the $\occ$ term for range reporting queries). We will need the following additional geometric data structures.
\begin{lemma}[J\'aJ\'a et al.~\cite{JMS2004}]\label{lem:2DRC}
For a set of $n$ points in $[0, u] \times [0,u]$ we can solve 2D range counting in $O(n)$ space, $O(n \log n)$ preprocessing time, and $O(\log n/\log \log n + \log \log u)$ query time. 
\end{lemma}
\begin{lemma}[van Emde Boas et al.~\cite{BKZ1977,Boas1977}, Mehlhorn and N\"aher~\cite{MN1990}]\label{lem:1DRC}
For a set of $n$ points in $[0,u]$ we can solve 1D range counting in $O(n)$ space, $O(n\log \log n)$ preprocessing time, and $O(\log \log u)$ query time. 
\end{lemma}
To achieve the result of Lemma~\ref{lem:1DRC} we use a van Emde Boas data structure~\cite{BKZ1977,Boas1977} implemented in linear space~\cite{MN1990} using perfect hashing. This data structure supports predecessor queries in $O(\log \log u)$ time. By also storing for each point it's rank in the sorted order of the points, we can compute a range counting query by two predecessor queries. To build the data structure efficiently we need to sort the points and build suitable perfect hash tables. We can sort deterministically in $O(n \log \log n)$ time~\cite{Han2004a}, and we can build the needed hash tables in $O(n)$ time using deterministic hashing~\cite{HMP2001} combined with a standard two-level approach (see e.g., Thorup~\cite{Thorup2003}).
\begin{lemma}[Chan et al.~\cite{CLP2011}]\label{lem:2DRE}
For a set of $n$ points in $[0, u] \times [0,u]$ we can solve 2D range emptiness in $O(n\log \log n)$ space, $O(n \log n)$ preprocessing time, and $O(\log \log u)$ query time. 
\end{lemma}

\subsection{The Data Structures}
We now show how to efficiently solve substring range counting and substring range emptiness. Recall that $S$ is a labeled string of length $n$ with labels from $[0,u]$. 

We can directly solve substring range counting by combining a suffix tree with the 2D range counting result from Lemma~\ref{lem:2DRC}. This leads to a solution using $O(n)$ space and $O(m + \log n/ \log \log n + \log \log u)$ query time. We show how to improve the query time to $O(m + \log \log u)$ at the cost of increasing the space to $O(n\log n/ \log \log n)$. Hence, we remove the $\log n/ \log \log n$ term from the query time at the cost of increasing the space by a $\log n / \log \log n$ factor. We cannot hope to achieve such a bound using a suffix tree combined with a 2D range counting data structure since any 2D range counting data structure using $O(n \log^{O(1)} n)$ space requires $\Omega(\log n /\log \log n)$ query time~\cite{Patrascu2007}. We can also directly solve substring range emptiness by combining a suffix tree with the 2D range emptiness result from Lemma~\ref{lem:2DRE}. This solution uses $O(n\log \log n)$ space and $O(m + \log \log u)$ query time. We show how to achieve optimal $O(m)$ query time with space $O(n\log \log u)$.

Our data structure for substring range counting and existence follows the construction in Section~\ref{sec:ds}. We partition the suffix tree into a top and a number of bottom trees and store a 1D data structure for each node in the top tree and a single 2D data structure. To answer a query for a pattern string $P$ of length $m$, we search the suffix tree with $P$ and use the 1D data structure if the search ends in the top tree and otherwise use the 2D data structure.

We describe the specific details for each problem. First we consider substring range counting. In this case the top tree consists of all nodes of string depth at most $\log n / \log \log n$. The 1D and 2D data structures used are the ones from Lemma~\ref{lem:1DRC} and~\ref{lem:2DRC}. By the same arguments as in Section~\ref{sec:ds} the total space used for the 1D data structures for all nodes in the top tree at depth $d$ is at most $O(n)$ and hence the total space for all 1D data structures is $O(n(\log n / \log \log n))$. Since the 2D data structure uses $O(n)$ space, the total space is $O(n\log n / \log \log n)$. The time to build all 1D data structures is $O(n(\log n / \log \log n) \cdot \log \log n))=  O(n\log n)$. Since the suffix tree and the 2D data structure can be built within the same bound, the total preprocessing time is $O(n\log n)$. Given a pattern of length $m$, a query uses $O(m + \log \log u)$ time if the search ends in the top tree, and $O(m + \log n/ \log \log n + \log \log u)$ time if the search ends in a bottom tree. Since bottom trees consists of nodes of string depth more than $\log n/ \log \log n$ the time to answer a query in both cases is $O(m + \log \log u)$. In summary, we have the following result.
\begin{theorem}\label{thm:counting}
Let $S$ be a labeled string of length $n$ with labels in the range $[0,u]$. We can solve substring range counting using $O(n\log n/\log \log n)$ space, $O(n\log n)$  preprocessing time, and $O(m + \log \log u)$ query time, for a pattern string of length $m$.
\end{theorem}

Next we consider substring range emptiness. In this case the top tree consists of all nodes of string depth at most $\log \log u$. We use the 1D and 2D data structures from Lemma~\ref{lem:1DRR} and Lemma~\ref{lem:2DRE}. The total space for all 1D data structures is $O(n \log \log u)$. Since the 2D data structure uses $O(n\log \log n)$ space the total space is $O(n\log \log u)$. For any constant $\gamma > 0$, the expected time to build all 1D data structures is $O(n \log \log u \log^{\gamma} u) = O(n \log^{\delta} u)$ for suitable constant $\delta >0$. The suffix tree and the 2D data structure can be built in $O(n \log n)$ time and hence the total expected preprocessing time is $O(n(\log n + \log^\delta u))$. If the search for a pattern string ends in the top tree the query time is $O(m)$ and if the search ends in a bottom tree the query time is $O(m + \log \log u)$. As above, the partition in top and bottom trees ensures that the query time in both cases is $O(m)$. In summary, we have the following result.
\begin{theorem}\label{thm:existence}
Let $S$ be a labeled string of length $n$ with labels in the range $[0,u]$. For any constant $\delta > 0$ we can solve substring range existence using $O(n \log \log u)$ space, $O(n(\log n + \log^\delta u))$ expected preprocessing time, and $O(m)$ query time, for a pattern string of length $m$. 
\end{theorem}

\section{Acknowledgments}
We thank Christian Worm Mortensen and Kasper Green Larsen for clarifications on the preprocessing times for the results in Lemma~\ref{lem:2DRC} and Lemma~\ref{lem:2DRE}.

\bibliographystyle{abbrv}
\bibliography{paper}

\end{document}